\input mnrass.sty
\pageoffset{-2.5pc}{0pc}
 

\Autonumber  


\pagerange{000--000}
\pubyear{2000}
\volume{000}

\begintopmatter  

\title{The distance to galactic globular clusters through RR Lyrae
  pulsational properties.}

\author{Cassisi S., De Santis R. \& Piersimoni A.M.}

\affiliation{Osservatorio Astronomico di Collurania, Via M. Maggini,
 I-64100, Teramo, Italy - E-Mail: cassisi,desantis,piersimoni@astrte.te.astro.it}

\shortauthor{Cassisi et al.}
\shorttitle{The GC distances by means of RR Lyrae properties.}


\acceptedline{}

\abstract
\tx

By adopting the same approach outlined in De Santis \& Cassisi
(1999), we evaluate the absolute bolometric magnitude of the Zero Age Horizontal
Branch (ZAHB) at the level of the RR Lyrae variable instability strip in
selected galactic globular clusters. This allows us to
estimate the ZAHB absolute visual magnitude for these clusters and to
investigate its dependence on the cluster metallicity. The derived $M_V(ZAHB) -
[Fe/H]$ relation, corrected in order to account for the luminosity difference 
between the ZAHB and the mean RR Lyrae magnitude, has been compared with 
some of the most recent empirical
determinations in this field, as the one provided by Baade-Wesselink
analyses, RR Lyrae periods, Hipparcos data for field variables and Main Sequence
fitting based on Hipparcos parallaxes for field subdwarfs. As a result,
our relation provides a clear support to the "long" distance scale.
We discuss also another method for measuring the distance to galactic globular
clusters. This method is quite similar to the one adopted for estimating the
absolute bolometric magnitude of the ZAHB but it relies only on the pulsational
properties of the Lyrae variables in each cluster.
The reliability and accuracy of this method has been tested by applying
it to a sample of globulars for which, due to the morphology of their horizontal
branch (HB), the use of the commonly adopted ZAHB fitting is a risky procedure.
We notice that the two approaches, for deriving the cluster distance modulus, provide
consistent results when applied to globular clusters whose RR Lyrae instability strip
is well populated.
Since the adopted method relies on theoretical predictions
on both the fundamental pulsational equation and the allowed mass range for
fundamental pulsators, we give an estimate of the error affecting present
results, due to systematic uncertainties in the adopted theoretical framework.


\maketitle  


\section{Introduction}
\tx 

RR Lyrae stars are the crossroad of several unsolved astrophysical
problems. 
In fact, even though thorough observational and theoretical investigations
have been devoted to analize the properties of these variables, we still
lack a firm understanding of systematics affecting both the slope and the 
zero-point affecting the $M_V(RR)$ vs. $[Fe/H]$ relation (Caputo 1997; 
Gratton 1998).
This unpleasant fact raised the so called
"distance dichotomy" between the "short" and the "long" distance scale.
In particular, we are dealing with the empirical evidence that both the
statistical parallaxes, and the Baade-Wesselink method of field RR Lyrae
stars seem to support the "short" distance scale, whereas the pulsation
properties of cluster RR Lyrae variables (Sandage 1993), the cluster
main-sequence fitting to local subdwarfs (Gratton et al. 1997), and the
calibration of HB luminosity based on the Large Magellanic Cloud distance 
obtained by adopting the Cepheid Period-Luminosity relation seem to support 
the "long" distance scale (Walker 1992).
On the other hand, distance determinations obtained by adopting both
evolutionary and pulsation predictions attain intermediate values
between the "short" and the "long" distance scale. In addition,
it has been suggested both by theoretical
and empirical evidences that the RR Lyrae Luminosity-Metallicity
relation - $M_V vs [Fe/H]$ - is not linear when moving from metal-poor to
metal-rich RR Lyrae stars (for a careful review on all these topics
we address to the reviews of Layden 1999, Popowski \& Gould 1999
and Gratton, Carretta \& Clementini 1999).

Up to now, a reliable estimate of the RR Lyrae luminosity -
metallicity relation has been hampered by several problems, mainly
related to the difficulties in estimating the individual cluster
distances, reddening and metallicity of both cluster and field RR
Lyrae variables.

This notwithstanding, the determination of the correct distance scale
for Population II stellar systems has a large impact on a wide range
of astrophysical problems, including the evaluation of the globular
cluster (GC) ages which provides a stringent constraint on the lower
limit to the age of the Universe, and the extra-galactic distances
measurement which is a fundamental step in deriving the Hubble constant.
Therefore, a precise determination of the absolute magnitude of the RR Lyrae
variables  - the traditional distance ladder for metal poor stellar
systems -, and of its dependence on the metallicity, is of crucial
importance.

In view of the remarkably different results obtained so
far with different methods, it is important to
analyze the problem by using as many as possible, different and
independent approaches in order to assess on more firm grounds
the $M_V - [Fe/H]$ relation for RR Lyrae stars.

In a previous paper (De Santis \& Cassisi 1999, hereinafter DC99), we
have adopted an approach based on the pulsational behavior of
RR Lyrae stars for obtaining an accurate estimate of the absolute
bolometric luminosity of ZAHB stars in GCs.

We notice that the results obtained by DC99 relied on the ranking 
of HB stellar mass as a function of the effective temperature predicted 
by evolutionary models. It is worth noticing that the effective temperature
of HB models, at variance with ZAHB luminosity, does not depend significantly
on the physical inputs adopted in the stellar computations.
As a consequence, it represents a rather reliable and 
trustworthy prediction of evolution theory. This occurrence allowed DC99
to perform a significant comparison with the most recent theoretical 
evaluations of the ZAHB luminosity (see Figure 15 in DC99).

In section 2 of this paper, we plan to use the results obtained in
DC99, in order to derive an accurate estimation of both the slope
and the zero point of the $M_V - [Fe/H]$ relation for ZAHB stellar
structures. In the same section, after applying a correction for the difference
between the mean RR Lyrae magnitude and the ZAHB one,
we check the consistence between our $<M_V>(RR Lyrae) - [Fe/H]$ relation
and the most recent empirical ones.
In section 3, we outline a method to determine the GCs distance based
on the pulsational properties of their RR Lyrae population.
The advantage of this method is that it does not need an estimate
of the ZAHB level: a step, which is particularly risky in the case of
GCs with blue HB, for which the RR Lyrae are suspected to be evolved
stars. A brief discussion and conclusions follow in the last section.

\section{\bf The $M_V(ZAHB)$ - [Fe/H] relation.}
\tx

The method employed for deriving the absolute bolometric luminosity
of ZAHB structure at a fixed effective temperature
(namely $\log{T_e}=3.85$) inside the RR Lyrae instability strip, as
well as the adopted temperature scale for the variables, have been
extensively discussed by DC99. Therefore, a detailed discussion
of the method will not be repeated here,
and we address the reader to the quoted paper for more details.

\figure{1}{S}{80mm}{\bf Figure 1. \rm {\sl Panel a}: The absolute visual ZAHB
magnitude versus the iron content for the GCs in our sample,
but NGC6171. The solid line represent our best fit to the observational
points obtained when excluding the data for NGC6171 (see text for more
details). {\sl Panel b}: Comparison between different solutions for
the mean absolute visual magnitude of RR Lyrae stars - metallicity
relation as provided by various authors (see labels).}

However, we wish to briefly recall the fundamental steps of
this approach. We make use of the results provided by the
updated pulsational models by Bono et al. (1997), about the relationship
among the fundamental period of the variable,
its mass, luminosity and effective temperature.
After rewriting the relation for the period as a function of
evolutionary properties of the variable
in order to obtain the dependence of the
pulsational reduced period ($P_{red}$) (Sandage 1981) on the mass and
effective temperature of the variable, and the absolute bolometric luminosity
of the ZAHB at $\log{T_e}=3.85$ ($L_{3.85}^{ZAHB}$), we have compared
in the $P_{red}$ - $T_{eff}$ diagram the
observational data for RR Lyrae stars with the theoretical prescriptions.

Since the theoretical relation for the reduced period depends on the
mass of the variable, we have used as lower and upper limit on this
mass the values provided by the evolutionary theory on the minimum and
maximum stellar mass which can produce a RR Lyrae star at a given fixed
metallicity. 
 
Our approach to estimate the stellar masses of RR Lyrae stars relies 
on predictions of nonlinear pulsational models, namely the edges of the 
instability strip as well as on evolutionary predictions concerning the 
ranking of stellar masses as a function of the effective temperature. 
As a consequence, it is worth discussing whether adopted pulsational 
and evolutionary predictions supply consistent evaluations for the 
RR Lyrae masses. Fortunately enough, Bono et al. (1996) in a recent 
investigation by adopting the same pulsational scenario we are adopting, 
showed that pulsational and evolutionary masses for RR Lyrae stars are 
in fair agreement.
Since, the evolutionary predictions (Castellani Chieffi \& Pulone 1991), 
adopted by Bono et al. (1996) to construct the pulsational models are 
quite similar to the evolutionary framework adopted in this investigation, 
namely the ranking of the stellar masses with effective temperatures, 
we are confident that pulsational and evolutionary predictions adopted 
in the current analysis are internally consistent. 
The reader interested in a more quantitative discussion concerning 
the difference of RR Lyrae masses provided by different evolutionary 
HB models is referred to DC99.

Then, for each cluster it has been quite easy to
determine the most suitable value for the intrinsic luminosity
of the ZAHB structures by properly fitting the lower and upper
boundaries of the RR Lyrae distribution in the reduced period -
temperature plane. It is worth remembering that this method allows
us to estimate the bolometric ZAHB luminosity with high accuracy,
and indeed the formal uncertainty on
$\log{L_{3.85}^{ZAHB}}$ is of the order of $\pm0.02$ (for an accurate
analysis of all uncertainties affecting the $L_{3.85}^{ZAHB}$
measurements we refer to DC99).

For clusters with also RRc (first overtone) variables
a similar approach has been adopted, obtaining results in fine agreement 
with the ones derived from RRab stars.

In order to successfully apply this method, one needs
clusters with homogeneous photometry for
both variable and non-variable HB stars, and spectroscopical
measurements of the metallicity. This strongly limits the size of
the sample of objects one can use. DC99 selected 7 clusters, namely:
NGC1851, NGC4590 (M68), NGC5272 (M3), NGC6171 (M107), NGC6362, NGC6981
(M72) and NGC7078 (M15). 
Data for other two clusters have been now added:
IC4499 (Walker \& Nemec 1996) and
NGC5904 (M5) (Caputo et al. 1999). For IC4499, we adopt the iron
abundance provided by Cohen et al. (1999), $\rm [Fe/H]=-1.46$,
whose metallicity scale is consistent with the one of
Carretta \& Gratton (1997, hereinafter CG97) adopted in DC99.
In the case of M5, we have adopted the iron content listed by CG97.

It is worth noticing that for this cluster, in order to test the
accuracy of our estimate, we have obtained the absolute bolometric
ZAHB luminosity by using also the larger sample of variables
investigated by Reid (1996). However, since Reid (1996) does not provide
the blue amplitude ($A_B$) needed for estimating the effective
temperature, we have derived $A_B$ from the visual amplitude ($A_V$),
by using the relation:
\smallskip
\line{$A_B=1.26\cdot{A_V}+0.04\cdot[Fe/H]+0.08\hfill 1)$}
\smallskip
\noindent
with a probable error p.e.=0.016 and a
correlation coefficient r=0.997, obtained from a sample of 12 field
variables observed by Liu \& James (1990). It is interesting to notice
that by using the two independent samples of variables, we have
obtained the same value for $L_{3.85}^{ZAHB}$.

Once obtained the value of $L_{3.85}^{ZAHB}$ for each cluster,
the absolute visual magnitude of the ZAHB can be easily derived
by using the relation:
\smallskip
\line{$M_V^{ZAHB}=-2.5\cdot\log{L_{3.85}^{ZAHB}}-(BC_{3.85}-BC_\odot)+
M_{V,\odot}\hfill 2)$}
\smallskip
\noindent
where $BC_{3.85}$ is the bolometric correction of a ZAHB star at
$\log{T_e}=3.85$. By using the bolometric correction scale by
Castelli, Gratton \& Kurucz (1997a,b), we have estimated that:
\smallskip
\line{$(BC_{3.85}-BC_\odot)=0.04[Fe/H]+0.14 \hfill 3)$}

\smallskip
\noindent
with a r.m.s. equal to 0.005 mag. For the absolute visual magnitude
of the Sun, we adopt $M_{V,\odot}=4.82\pm0.02$ mag (Hayes 1985).

For all clusters in our sample, Table 1 reports the most relevant
quantities: the name of the cluster, the adopted visual magnitude of
the ZAHB, the iron abundance, the absolute bolometric luminosity of
the ZAHB at $\log{T_e}=3.85$, the absolute visual ZAHB magnitude,
the distance modulus and the reddening. For the clusters with
both fundamental and first overtone variables, the result listed
has been obtained by averaging the values estimated from the RRab
and RRc variables. The reddening values have been derived by comparing
in the $(B-V) - A_B$ plane, the observational data for the
RRab Lyrae stars with the empirical relation derived by
Caputo \& De Santis (1992).
 
It is worth investigating how much our predicted ZAHB absolute 
visual magnitudes are affected by systematic uncertainties in the main 
ingredients adopted in the method previously discussed. In DC99, we showed 
that when moving from the old pulsation relation,
i.e. the relation connecting the period, stellar mass, luminosity, and 
effective temperature, provided by van Albada \& Baker (1971) to the recent 
relation by Bono et al. (1997) the difference in the predicted $M_V^{ZAHB}$ is 
quite small and roughly equal to 0.04 mag. 
A further pulsational input we are adopting is the position in the HR diagram 
of the fundamental instability strip. Recently, Caputo et al. (2000) have 
investigated the dependence of the First Overtone Blue Edge (FOBE) on the 
Helium abundance and the uncertainty on the treatment of superadiabatic 
convection. They have found that the location of the instability strip is 
marginally affected by mild He variations. Moreover, they also tested that 
current uncertainties in the calibration of the mixing length parameter 
cause a change (see discussion in Caputo et al. 2000) in the period at 
the FOBE at most of the order of $\Delta\log{P}=\pm0.03$. This difference 
implies an uncertainty in the temperature of the FOBE of 
$\Delta\log{T_e}(FOBE)=\pm 0.009$. 
If we assume that this shift affects simultaneously the blue and the 
red edge of fundamental and first overtone pulsators then the maximum 
estimated error on the fundamental RR Lyrae masses is smaller than $0.005M_\odot$. 
The impact of this uncertainty on predicted $M_V^{ZAHB}$ is negligible and roughly 
equal to 0.005 mag.
As far as the uncertainty of current RR Lyrae temperature scale 
is concerned, DC99 emphasized that the 
probable error affecting the temperature estimate for each variable 
is equal to $\Delta\log{T_e}=\pm0.003$. If we assume that the temperature determinations
of the entire RR Lyrae sample in a cluster are systematically affected by 
this uncertainty, then the predicted  ZAHB absolute visual magnitude is  
affected by an error of the order of 0.03 mag. As a whole, by accounting 
for all the previous error sources, we find that current $M_V^{ZAHB}$ values 
could be affected, at most, by uncertainties of the order of 0.05 mag.

The values of $M_V^{ZAHB}$ for all clusters, but NGC6171 (see below),
are shown in Figure 1a. As far as it concerns the uncertainty on
[Fe/H], we account for a realistic indetermination of about 0.15 dex
(see Rutledge, Hesser \& Stetson 1997). By using the data plotted
in this Figure and listed in Table 1, we can now derive the
$M_V - [Fe/H]$ relation by performing a best fit of the observational
points.
Since DC99 have shown that the value of $L_{3.85}^{ZAHB}$
for NGC6171 is affected by a large uncertainty, due to the poor quality
of the available photometry, it has not been taken into
account in order not to bias the solution.

By accounting for the uncertainty on both the absolute visual
magnitude and the metallicity we derive the following relation:
\smallskip
\line{$M_V^{ZAHB}= (0.17\pm0.03)\cdot[Fe/H]+(0.87\pm0.04) \hfill 4)$}
\smallskip
\noindent

\figure{2}{D}{120mm}{\bf Figure 2. \rm Comparison in the
$(\log{P}+0.33\cdot{<V>}) - \log{T_e}$ diagram between the $RR_{ab}$
variables in different GCs and the prescriptions provided by the
equation 5), when fixing the GC distance modulus to the value listed in Table 2,
and adopting for the allowed minimum and maximum variable mass the
values provided by the stellar evolutionary theory. Temperature scale
is from De Santis (1996).}

In view of the significant uncertainties still affecting the GCs
metallicity scale, we have decided to perform the
$M_V(ZAHB) - [Fe/H]$ calibration by using also the Zinn \& West (1984)
metallicity scale. The value of $L_{3.85}^{ZAHB}$
for each cluster in our sample has been recomputed by using the
Zinn \& West (1984)'s scale, and the final calibration is the following:
\smallskip
\line{$M_V^{ZAHB}= (0.17\pm0.03)\cdot[Fe/H]+(0.88\pm0.05) \hfill 5)$}
\smallskip
\noindent
which is in fine agreement with the result based on the CG97 scale.

Since in the literature, different relations between the absolute
magnitude of HB stars and the heavy elements abundance can be found,
we have decided to compare the most recent ones with our solution.
This has been done in fig. 1b. Since in DC99, we have already
compared our determinations of $L_{3.85}^{ZAHB}$ with the most
significant theoretical evaluations, now we limit the comparison to
the empirical determinations of the $M_V(RR) - [Fe/H]$ relation.
More in detail, we take into account the solutions given by
Walker (1992), Sandage (1993), Clementini et al. (1995), Feast (1997),
Gratton et al. (1997), Fernley et al. (1998), Groenewegen \& Salaris
(1999) and Caputo et al. (2000).

Since all these relations refer to the mean magnitude of the RR Lyrae,
we have applied a correction to our solution in order to obtain the
RR Lyrae mean magnitude from the ZAHB luminosity level:
\smallskip
\line{$<V_{RR}>=V_{ZAHB}-0.04\cdot[Fe/H]-0.15 \hfill 6)$}
\smallskip
\noindent
provided by Cassisi \& Salaris (1997, but see also Carney et al. 1992).

From data in fig. 1b), one can easily notice that this
relations is in satisfactory agreement (within $\approx0.07$ mag) for
$[Fe/H]<-1.5$ with the corresponding relation obtained by Caputo et al. (2000) 
in the same metallicity range, when assuming a solar-scaled distribution for 
the heavy elements. However, we are not able to assess the existence of a change
in the slope of the $M_V(RR) - [Fe/H]$ relation as disclosed by the quoted
authors due to small size of the cluster sample, and in particular to the lack
of GCs in the relevant metallicity range: $-2.0<[Fe/H]<-1.5$. 
As a consequence, the two solutions differ also by about
$\approx0.10$ mag at the upper metallicity limit ($[Fe/H]\approx-1$)
we explore.

\table{1}{D}{\bf Table 1. \rm The main properties of the selected sample of
globular clusters.}
{\halign{%
\rm#\hfil& \hskip7pt\hfil\rm#\hfil&\hskip7pt\hfil\rm#\hfil&
\hskip7pt\hfil\rm#\hfil&\hskip7pt\hfil\rm#
\hfil& \hskip7pt\hfil\rm#\hfil & \hskip7pt\hfil\rm#\hfil &
\hskip7pt\hfil\rm#\hfil & \hskip7pt\hfil\rm\hfil#
\hfil & \hskip7pt\hfil\rm#\hfil & \hskip7pt\hfil\rm#\hfil\cr
NGC & Name &  $V_{ZAHB}$ &[Fe/H] & $\log{L_{3.85}^{ZAHB}}$ & $M_{V}^{ZAHB}$ & $(m-M)_V$ & $E(B-V)$ \cr
\noalign{\vskip 10pt}
6171  & M107 & 15.85$\pm$0.10 & -0.87  & 1.54$\pm$0.04  & 0.874$\pm$0.10 & 14.98$\pm$0.02 & 0.38$\pm$0.02\cr
6362  &    & 15.33$\pm$0.03 & -0.96  & 1.615$\pm$0.015& 0.693$\pm$0.03 & 14.64$\pm$0.01 & 0.07$\pm$0.013 \cr
1851  &    & 16.13$\pm$0.025& -1.08  & 1.645$\pm$0.01 & 0.618$\pm$0.025& 15.51$\pm$0.01 & 0.045$\pm$0.016\cr
5904  & M5 & 15.20$\pm$0.05 & -1.11  & 1.60$\pm$0.03  & 0.733$\pm$0.08 & 14.47$\pm$0.03 & 0.06$\pm$0.02  \cr
6981  & M72& 17.07$\pm$0.03 & -1.30  & 1.635$\pm$0.01 & 0.653$\pm$0.03 & 16.42$\pm$0.01 & 0.07$\pm$0.015 \cr
5272  & M3 & 15.73$\pm$0.02 & -1.34  & 1.645$\pm$0.01 & 0.628$\pm$0.02 & 15.10$\pm$0.01 & 0.012$\pm$0.01 \cr
      & IC4499& 17.72$\pm$0.05 & -1.46  & 1.65$\pm$0.03  & 0.619$\pm$0.08 & 17.10$\pm$0.05 & 0.21$\pm$0.01 \cr
4590  & M68& 15.70$\pm$0.02 & -1.99  & 1.70$\pm$0.02  & 0.506$\pm$0.02 & 15.19$\pm$0.03 & 0.043$\pm$0.014\cr
7078  & M15& 15.92$\pm$0.05 & -2.12  & 1.68$\pm$0.02  & 0.556$\pm$0.05 & 15.36$\pm$0.03 & 0.074$\pm$0.014\cr}}


It seems also to exist a satisfactory agreement (at the level of less
than 0.1 mag) with the relations given by Walker (1992),  Groenewegen \& Salaris (1999)
and by Gratton et al. (1997). Therefore, present result provides further
support to the "long" distance scale. On the contrary, an evident
disagreement exists  with the results based on the Baade-Wesselink
method, as those provided by Clementini et al. (1995), Fernley et al. (1998)
and Feast (1997).

\section{\bf A pulsational approach to the GCs distance.}
\tx

In Table 1, we have reported for each cluster
the apparent distance modulus as obtained by using the
$V_{ZAHB}$ estimates following 
DC99, and the value of $M_V^{ZAHB}$ obtained in the previous section.
DC99 have already shown that the major source of uncertainty in
the estimation of $L_{3.85}^{ZAHB}$ (and, in turn, of $M_V^{ZAHB}$)
relies in the evaluation of the apparent magnitude of the ZAHB (see also
the discussion in the previous section).
However, it is worth emphasizing that, due to the approach adopted for
deriving $M_V^{ZAHB}$, the measurement of the distance modulus
$(m-M)_V=V_{ZAHB}-M_V^{ZAHB}$ is no more affected by any possible
uncertainty in the choice of the ZAHB level.

This is a quite important point, since it means that the main source
of uncertainty in the measurement of the distance modulus is related
to the indetermination - usually very small - on $L_{3.85}^{ZAHB}$
due to the fit procedure in the period - effective temperature
diagram between theory and observations. As a consequence, this
method allows us to determine the distance modulus of clusters with
a rich population of variables and spectroscopical measurements of
their metallicity, with an uncertainty usually very small ($\le0.03$
mag).

In passing, we wish to notice that our distance modulus estimation for M3 appears
in fine agreement with the distance recently derived by Bono et al. (2001)
by adopting an independent method namely the K-band period-luminosity 
relation of RR Lyrae ($15.03\pm0.07$ mag).

In the following, we wish to outline a method useful for measuring the
distance modulus of galactic GCs, based only on the pulsational
properties of their RR Lyrae population. This method appears
particularly attractive in case of GCs, showing a very blue HB (usually the ones with
HB type larger than 0.8 ). In fact, in these clusters, it is no more
possible to identify the lower envelope of the observed HB as the ZAHB
locus. 
In fact, many (if not all) stars, within the RR Lyrae instability strip, are 
significantly evolved objects, crossing the strip at magnitudes brighter than
the ZAHB level, during their evolution toward the Asymptotic Giant Branch.

Besides, we notice that, whereas for determining
the $L_{3.85}^{ZAHB}$ value for each cluster, one must
rely on homogeneous photometry for both variable and non-variable HB
stars (DC99), for obtaining the distance modulus by using the
following approach, one needs to know only the mean magnitudes and the
pulsational  properties of RR Lyrae variables.

The suggested method works as follows: for each cluster variable 
we implement eq. 2) in DC99 with eq. 2) and 3) of the present
paper, and the fundamental pulsational equation can be easily written in the
following form:
\smallskip
\line{$\log{P}+0.33\cdot{<V}>=0.33\cdot(m-M)_V-0.33\cdot\Delta{BC}+$\hfill}

\line{\hskip 0.4truecm
$-0.013\cdot[Fe/H]-0.582\log{M}-3.506\log{T_e}+13.171 \hfill 6)$}
\smallskip
\noindent
where $<V>$ is the mean visual magnitude, $(m-M)_V$ is the apparent
distance modulus of the cluster,
$\Delta{BC}=BC_{\log{T_e}}-BC_{3.85}=-5.252(\log{T_e})^2+
41.636\log{T_e}-82.454$
i.e. the difference between the bolometric correction of a HB
structure in the instability at $\log{T_e}=3.85$ and the one with
the same effective temperature of the variable (see DC99); the
other quantities have their usual meaning.
By using the same approach as in DC99, it is evident that once fixed
the cluster metallicity and the allowed mass range for the RR Lyrae
variables, one can use the $(\log{P}+0.33\cdot{<V>}) - \log{T_e}$
diagram for constraining the $(m-M)_V$ value.

Concerning the evaluation of the minimum and maximum mass for
fundamental pulsators, we follow the approach suggested by DC99, which
is based on the determination, for each metallicity, of the structures 
spending within the instability strip a significant amount ($\approx20\%$) 
of their whole core He-burning phase. 
This approach relies on evolutionary lifetimes within the instability strip, 
and does not account for individual cluster HB morphology. However, synthetic 
HB experiments disclose that our approach safely estimates the mass range 
of variables in metal-poor clusters. On the other hand, in the case of 
intermediate metallicity clusters, affected by the 2nd parameter effect, 
the simulations show that our method safely estimates the minimum stellar 
mass which produces RR Lyrae stars, but slightly overestimate the maximum 
pulsator mass.
Nevertheless, due to the dependence of $(\log{P}+0.33\cdot{<V>})$ on the
variable mass and to the fit procedure between theory and observations, 
it results that an error on the upper mass limit of the order of $\approx0.05M_\odot$, 
causes a change in the estimation of the GC distance modulus of the order of 
$\approx0.02$ mag.

In order to show better how this method works, we have applied it to a
selected sample of clusters. In particular, we have chosen the following
GCs: M92, NGC6426, NGC5053, NGC5466, M55, M9 and M2.
One has to notice that for the clusters M9, M55 and NGC 6426 only
the $RR_{ab}$ V amplitude are available. Since the adopted pulsational
temperature scale (De Santis 1996) has been calibrated on blue
amplitude, we used eq. 1) to obtain the blue amplitude for each variable
in the sample.

In the case of the cluster NGC6426, we have not accounted for one
(variable V16) out of the 8 variables investigated by Papadakis et al.(2000), as
the classification of this variable is ambiguous, being suspected to be a c
type variable. In the case of M9, we have omitted the variable V7 because
its apparent magnitude is affected from an obscuring cloud to the
southwest of the cluster.

We wish to notice that all these clusters are characterized by blue HB
morphology. In Table 2, we list for each cluster in this sample, the
reference for the RR Lyrae data, the adopted iron abundance as provided 
by CG97, the HB type defined as (B-R)/(B+V+R) -where B,V and R are the numbers 
of stars hotter than the RR Lyrae instability strip, of RR variables, 
and of stars cooler than the instability strip- provided by Harris (1996),
the estimated mass range for fundamental pulsators and the distance modulus 
obtained by using the previously described approach and the related uncertainty.

In Figure 2, for each cluster we show the comparison between theory and
observations in the $(\log{P}+0.33\cdot{<V>}) - \log{T_e}$ diagram. 
One can notice that, once fixed the minimum and maximum allowed mass 
for fundamental pulsators, the observational distribution is well matched 
only when fixing the GC distance modulus to the value listed in Table 2.

\table{2}{D}{\bf Table 2. \rm The main properties of the GC sample to which
has been applied our method for measuring the distance modulus.}
{\halign{%
\rm#\hfil& \hskip7pt\hfil\rm#\hfil&\hskip7pt\hfil\rm#\hfil& \hskip7pt\hfil\rm#\hfil&\hskip7pt\hfil\rm#
\hfil& \hskip7pt\hfil\rm#\hfil &  \hskip7pt\hfil\rm\hfil# \hfil &
\hskip7pt\hfil\rm#\hfil \cr
NGC & Name & Source & [Fe/H]  & HB Type & $M_{RR}/M_\odot$ & $(m-M)_V$\cr
\noalign{\vskip 10pt}
 6341 & M92 & Carney et al. (1992)     & -2.15 & 0.90 & 0.70-0.80 & 14.71$\pm$0.05\cr
 6426 &     & Papadakis et al. (2000)  & -2.07 & 0.58 & 0.70-0.80 & 17.80$\pm$0.07\cr
 5053 &     & Nemec et al. (1995)      & -2.10 & 0.52 & 0.70-0.80 & 16.13$\pm$0.05\cr
 5466 &     & Corwin et al. (1999)     & -2.03 & 0.58 & 0.70-0.80 & 16.05$\pm$0.03\cr
 6809 & M55 & Olech et al. (1999)       & -1.65 & 0.87 & 0.68-0.74 & 13.90$\pm$0.05\cr
 6333 & M9  & Clement \& Shelton (1999)& -1.57 & 0.87 & 0.67-0.73 & 15.75$\pm$0.05\cr
 7089 & M2  & Lee \& Carney (1999)     & -1.34 & 0.96 & 0.64-0.70 & 15.61$\pm$0.05\cr}}

The estimate of the M92 distance modulus appears in satisfactory
agreement with the results provided by Pont et al. (1998), Carretta et al. (2000)
and Vandenberg (2000); and within the listed uncertainties also with the estimation given
by Reid \& Gizis (1998). As far as the GCs in common with the work of Caputo et al. (2000),
we verify a general agreement within 0.1 mag. However, for two clusters,
namely NGC5053 and M2, it exists a large discrepancy of 0.18 and 0.16
mag respectively.

We wish to notice that this method, when applied to GCs whose RR Lyrae 
instability strip is well populated, as for instance the clusters listed in Table 1,
provides distance determinations fully consistent with the ones obtained by using the
approach adopted in the previous section for deriving 
$L_{3.85}^{ZAHB}$, and in turn, $M_V^{ZAHB}$.

\section{\bf Conclusions.}
\tx

In present work, by adopting the same approach discussed in DC99, we
increase the sample of galactic GCs for which we derive
the absolute bolometric magnitude of the ZAHB ($L_{3.85}^{ZAHB}$) at the
level of the RR Lyrae instability strip. As already shown in DC99, this
occurrence is quite important in order to test the accuracy and reliability of the
current theoretical predictions on this quantity.
In addition, we now use the obtained values for $L_{3.85}^{ZAHB}$ in
order to estimate, for each cluster in our sample, the absolute visual magnitude of the
ZAHB. This allows us to investigate the dependence of this parameter on the cluster
metallicity, by deriving a $M_V(ZAHB) - [Fe/H]$ relation. Due to the
limited number of GCs in our sample, we can not assess, as made by Caputo et al. (2000), 
the existence of a non-linear dependence of $M_V^{ZAHB}$ on the cluster metallicity.

The comparison of our $M_V(ZAHB) - [Fe/H]$ relation - after rescaling it
for the luminosity difference between the ZAHB and the mean RR Lyrae magnitude -, 
with the most recent empirical determinations of $M_V(RR) - [Fe/H]$ relation, 
shows that our result is in satisfactory agreement with almost all measurements 
supporting the "long" distance scale as the ones provided by Gratton et al. (1997), 
Groenewegen \& Salaris (1999) and Walker (1992).
Present result is also in good agreement with the Caputo et al. (2000)
determination for metallicity lower than -1.5 dex. For larger metallicity, since the
relation by Caputo et al. (2000) has a stronger dependence on metallicity than our one, 
it exists a significant discrepancy, increasing with the metallicity, also of 
the order of 0.1 mag. 
A clear discrepancy appears between our distance scale and all the
other ones supporting the "short" distance scale as the ones based on
Baade-Wesselink method (Clementini et al. 1995, Fernley et al. 1998).

We present a method for determining the distance to galactic GCs, based
only on the pulsational properties of RR Lyrae stars. This method consists in
comparing observations and expectations provided by updated pulsational and
evolutionary models in the $(\log{P}+0.33\cdot{<V>}) - \log{T_e}$ diagram.

From a theoretical point of view, our approach relies on predictions like the 
dependence of the fundamental pulsational equation on the evolutionary properties 
of the variable, and the allowed mass range for fundamental pulsators, therefore
the accuracy of the obtained results depends on the reliability of the adopted
theoretical framework. However, since current theoretical predictions on both
the fundamental pulsation equation and the ranking of HB stellar mass as a 
function of the effective temperature, appear quite robust, we think that this method 
can provide accurate distance determinations.
Nevertheless, it does also need the use of a temperature scale for RR
Lyrae stars, whose reliability is a long-standing problem (see, for instance, Catelan (1998) and 
Carretta, Gratton \& Clementini (2000)).
However, in DC99 (see also De Santis 2001) we have carefully checked the accuracy
of the temperature scale provided by De Santis (1996) and adopted in the
present investigation.
So, we are confident that our distance modulus determinations are not
significantly affected by the residual uncertainty affecting the RR Lyrae temperature
scale.

The reliability of the suggested method has been shown by deriving the
distance moduli of a selected sample of galactic GCs, for which the determination of the
distance through the usual HB fitting has been always a risky procedure due to
the morphology of their HB. The derived distances appear, within the
uncertainty, in satisfactory agreement with the values listed in the more recent literature.

\section*{Acknowledgments}
\tx 

We wish to warmly thank G. Bono, for all interesting and stimulating suggestions 
and for the constant and generous help provided all along these years.
It is a real pleasure to thank F. Caputo for a detailed reading of an early draft
of this paper, as well as for several interesting discussions on this topic.
We wish also to warmly thank M. Salaris for an accurate reading of a preliminary draft
and for his suggestions. We sincerely acknowledge M. Corwin, B. Carney and  J.W. Lee 
for kindly supplying their data for the RR Lyrae stars in NGC5466 and M2. 
We warmly thank the referee R. Gratton for the detailed reading 
of our manuscript and for the pertinence of his comments which have significantly
improved the content and the readability of the paper.
This work was supported by the Ministero Italiano dell'Universit\'a e della Ricerca 
Scientifica e Tecnologica (MURST) (Cofin2000) under the scientific project "Stellar 
observables of cosmological relevance".

\section*{References}
\bibitem Bono, G., Caputo, F., Castellani, V. \& Marconi, M. 1996, ApJL, 471, 33
\bibitem Bono, G., Caputo, F., Castellani, V. \& Marconi, M. 1997,
A\&AS, 121, 327
\bibitem Bono, G., Caputo, F., Castellani, V., Marconi, M., \& Storm, J. 2001,
MNRAS, {\sl submitted to}
\bibitem Caputo, F. 1997, MNRAS, 284, 994
\bibitem Caputo, F., Castellani, V., Marconi, M., \& Ripepi, V. 1999,
MNRAS, 306, 815
\bibitem Caputo, F., Castellani, V., Marconi, M., \& Ripepi, V. 2000,
MNRAS, 306, 819
\bibitem Caputo, F. \& De Santis R. 1992, AJ, 104, 253
\bibitem Carney B.W., Storm, J., Trammell, S.R. \& Jones, R.V. 1992,
PASP, 104, 44
\bibitem Carretta, E. \& Gratton, R.G. 1997, A\&AS, 121, 95
\bibitem Carretta, E., Gratton, R.G. \& Clementini,G. 2000, MNRAS, 316, 721
\bibitem Carretta, E., Gratton, R.G., Clementini,G. \& Fusi Pecci F. 2000, ApJ, 533, 215
\bibitem Cassisi, S. \& Salaris, M. 1997, MNRAS, 285, 593
\bibitem Castellani, V., Chieffi, A., \& Pulone, L. 1991, ApJS, 76, 911 
\bibitem Castelli, F., Gratton, R.G. \& Kurucz, R.L. 1997a, 318, 841
\bibitem Castelli, F., Gratton, R.G. \& Kurucz, R.L. 1997b, 324, 432
\bibitem Catelan, M. 1998, ApJ, 495, L81
\bibitem Clement, C.M. \& Shelton, I., 1999, AJ, 118, 453
\bibitem Clementini, G., Carretta, E., Gratton, R., Merighi, R., Mould,
J.R., \& McCarthy, J.K. 1995, AJ, 110, 2319
\bibitem Cohen, J.G., Gratton, R.G., Behr, B.B., \& Carretta, E. 1999, ApJ, 523, 739
\bibitem Corwin,T.M.,Carney, B.W. \& Nifong, B.G. 1999, AJ, 118, 2875
\bibitem De Santis, R. 1996, A\&A, 306, 755
\bibitem De Santis, R. 2001, MNRAS, {\sl submitted to}
\bibitem De Santis, R. \& Cassisi, S. 1999, MNRAS, 308, 97 (DC99)
\bibitem Feast, M.W. 1997, MNRAS, 284, 761
\bibitem Fernley, J., Barnes, T.G., Skillen, I., Hawley, S.L., Hanley,
C.J., Evans, D.W., Solano, E. \& Garrido, R. 1998, A\&A, 330, 515
\bibitem Gratton, R.G. 1998, MNRAS, 296, 739
\bibitem Gratton, R.G., Carretta, E. \& Clementini, G. 1999,
in "Post-Hipparcos cosmic candles", Kluwer Acad. Publ., A. Heck \&
F. Caputo eds., p. 89
\bibitem Gratton, R.G., Fusi Pecci, F., Carretta, E., Clementini, G.,
Corsi, C.E., \& Lattanzi, M. 1997, ApJ, 491, 749
\bibitem Groenewegen, M.A.T., \& Salaris, M. 1999, A\&A, 348, L33
\bibitem Harris, W.E. 1996, AJ, 112, 1487
\bibitem Hayes, D.S. 1985, IAU Symp. 111, p.225
\bibitem Layden, A.C. 1999, in "Post-Hipparcos cosmic candles",
Kluwer Acad. Publ., A. Heck \& F. Caputo eds., p. 37
\bibitem Lee, J.W. \& Carney, B.W., 1999, AJ, 117, 2868
\bibitem Liu, T. \& James, K.A. 1990, ApJ, 354, L273
\bibitem Nemec, J.M., Mateo, M. \& Schombert, J.M., 1995, AJ, 109, 618
\bibitem Olech, A., Kaluzny, J., Thompson, I.B., Pych, W., Krzeminski,
W. \& Shwarzenberg-Czerny, A., 1999, AJ, 118, 442
\bibitem Papadakis, I., Hatzidimitriou, D., Croke, B.F.W. \&
Papamastorakis, I. 2000, AJ, 119, 851
\bibitem Pont, F., Mayor, M., Turon, C., \& Vandenberg, D.A. 1998,
A\&A, 329, 87
\bibitem Popowski, P. \& Gould, A. 1999, in "Post-Hipparcos cosmic
candles", Kluwer Acad. Publ., A. Heck \& F. Caputo eds., p. 53
\bibitem Reid, I.N. 1996, MNRAS, 278, 367
\bibitem Reid, I.N., \& Gizis, J.E. 1998, AJ, 116, 2929
\bibitem Rutledge, G. A., Hesser, J. E., \& Stetson, P. B. 1997, AJ, 109, 907
\bibitem Sandage, A. 1981, ApJ, 248, 161
\bibitem Sandage, A. 1993, AJ, 106, 719
\bibitem van Albada, T.S., \& Baker, N. 1971, ApJ, 169, 311
\bibitem Vandenberg, D.A. 2000, ApJS, 129, 315
\bibitem Walker, A.R. 1992, ApJ, 390, L81
\bibitem Walker, A.R., \& Nemec, J.M.  1996, AJ, 112, 2026
\bibitem Zinn, R. \& West, M.J. 1984, ApJS, 55, 45
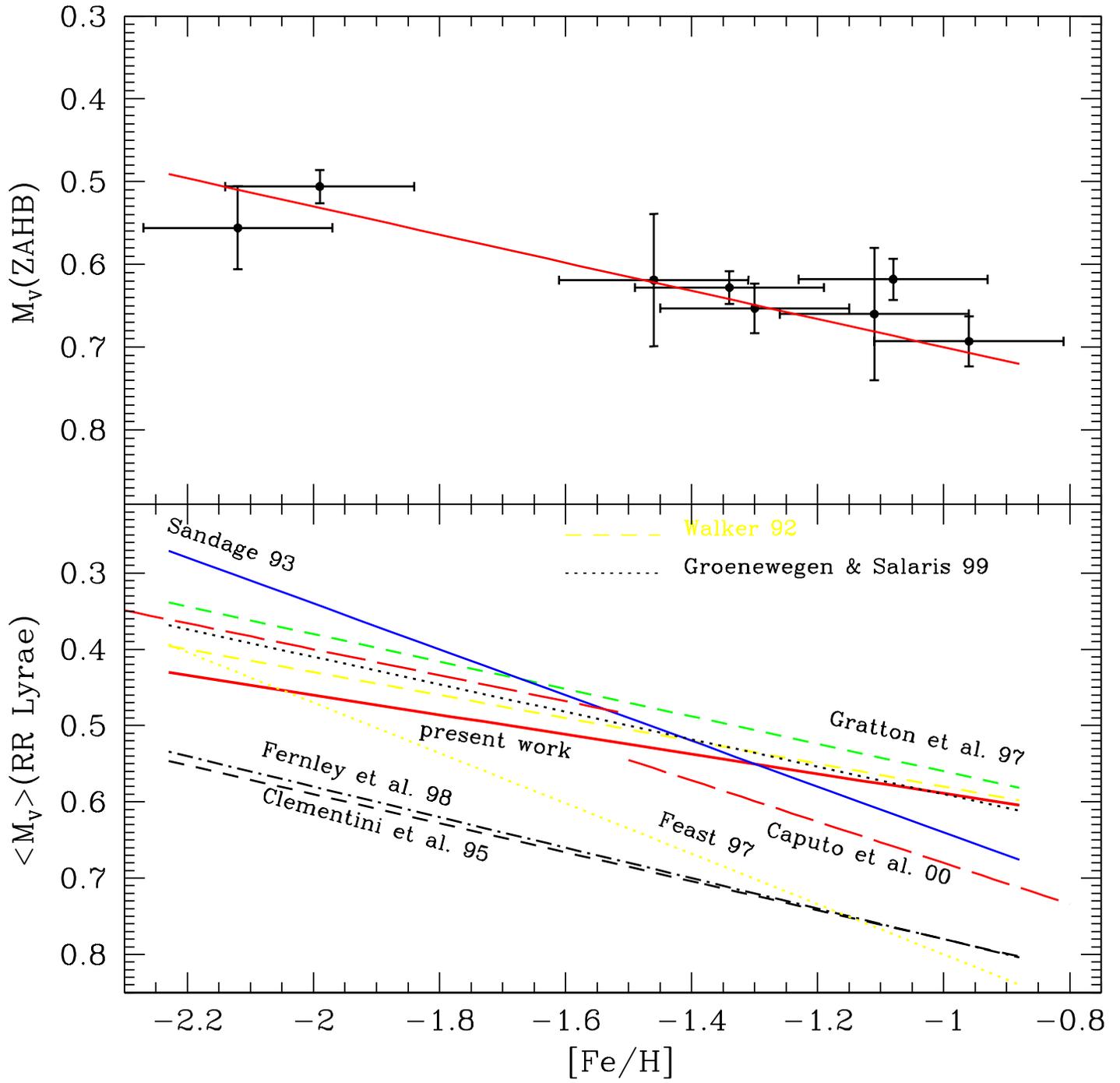
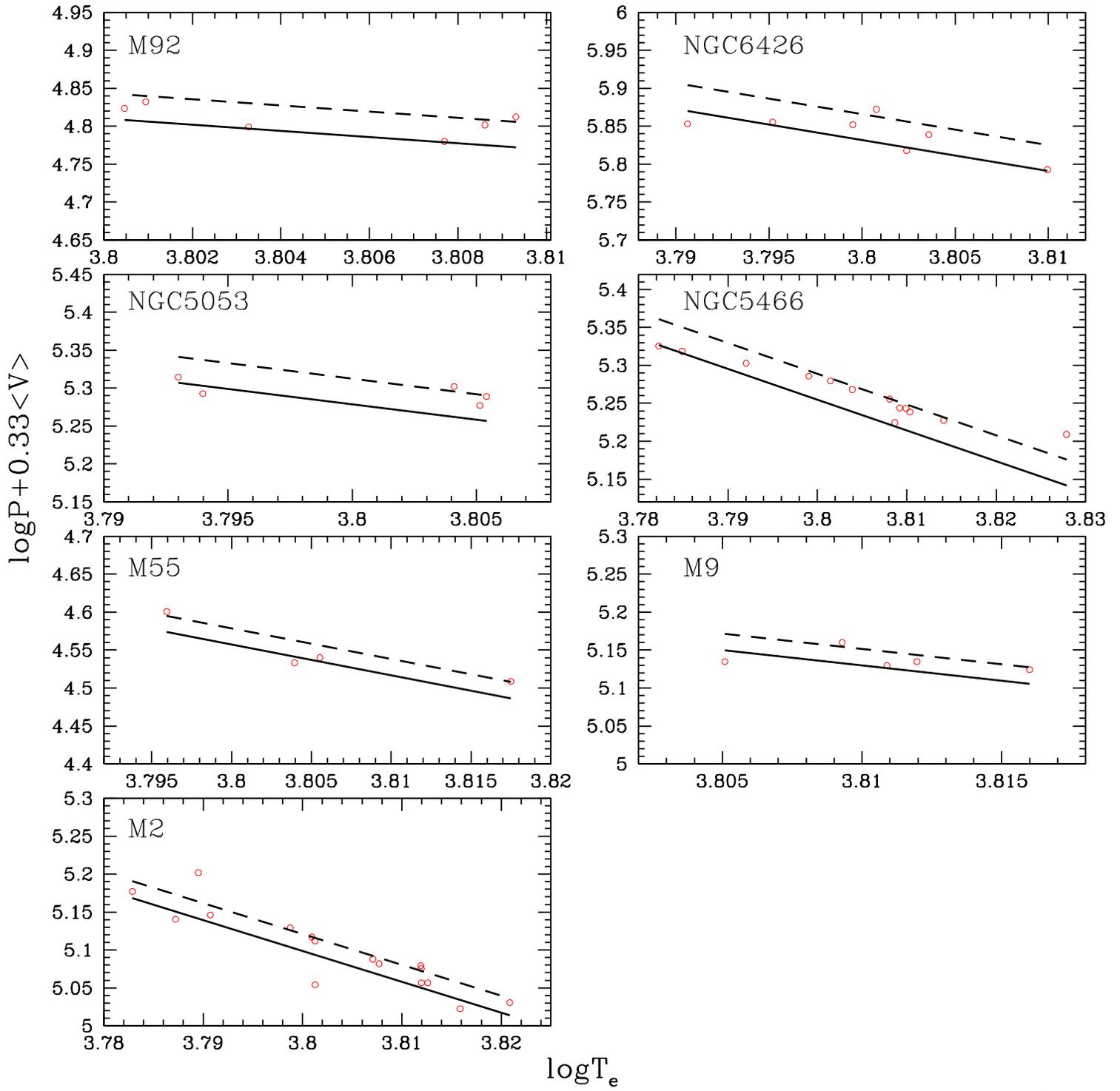
\bye